# Transferability of force fields for 2D silicon (silicene)

Marcin Maździarz[*]

Address: Department of Computational Science, Institute of Fundamental Technological Research Polish Academy of Sciences, Pawińskiego 5B, 02-106 Warsaw, Poland

Email: Marcin Maździarz - mmazdz@ippt.pan.pl

[*] Corresponding author

## Abstract

An ability of various interatomic potentials to reproduce the properties of silicene (2D silicon) polymorphs were examined. Structural and mechanical properties of the flat (FS), low-buckled (LBS), trigonal dumbbell (TDS), honeycomb dumbbell (HDS) and large honeycomb dumbbell (LHDS) single-layer silicon (silicene) phases, were obtained using density functional theory (DFT) and molecular statics (MS) calculations with Tersoff, MEAM, Stillinger-Weber, EDIP, ReaxFF, COMB and machine-learning-based (ML-IAP) interatomic potentials. A quantitative systematic comparison and discussion of the results obtained are reported.

## Keywords

2D materials; Silicene; Interatomic potentials; Force fields; DFT; Mechanical properties

# 1 Introduction

We live in the "Silicon Age" due to the great importance of elemental silicon to the modern global economy, it is mainly about electronics. Silicon is one of the best investigated materials and the quality of its production is impressive. However, this applies to bulk silicon. The interest in the graphene has also led to interest in other non-carbon 2D materials [1,2]. One such material is precisely the 2D silicon called silicene [3,4]. Using first-principles methods, with current com-puter resources, we can model structures up to about a few hundred atoms. For larger systems, we need approximate methods, e.g. molecular dynamics/statics. For these methods, the qual-ity of the interatomic potentials (IAPs) used is crucial. Because of the importance of silicon, but also its complexity, dozens of potentials have been proposed for it. In the very well-known NIST Interatomic Potentials Repository, there are 27 potentials for silicon (no other element has more potentials), the oldest one from 1985 and the latest from 2020 [5]. At least five 2D silicon polymorphs have been reported in the literature, that is, flat (FS), low-buckled (LBS) [6], trigonal dumbbell (TDS), honeycomb dumbbell (HDS) and large honeycomb dumbbell (LHDS) [7]. There are still doubts about their dynamic stability, e.g. for a flat phase the negative ZO phonon mode could be removed by the selection of an appropriate substrate, see [3,8]. The ability of potentials for 3D silicon to reproduce 2D silicon is poorly studied. There are several papers where the quality of potentials for 3D silicon has been assessed, see e.g. [9-11] but not for silicene. The intention of this work is first to determine the structural and mechanical properties of 2D silicon using first-principles method and then to test the ability of different interatomic poten-tials to reproduce these properties.

## 2 Methods

Analyzing the available literature concerning all phases of single-layer (SL) Si, it is not feasible to find all structural, mechanical and phonon data obtained in one consistent way. The availabil-ity of experimental data is actually limited to the silicene grown on some support, a pristine free-standing SL sheet of silicene has not been discovered to date [4,12], and therefore we must use *ab initio* calculations. Unfortunately, also *ab initio* calculations, most often DFT, differ in the calcula-tion methodology, i.e., they use different functional bases, different pseudopotentials or exchange-correlation (XC) functionals, and parameters such as cohesive energy and elastic constants are poorly accessible. For this reason, structural and mechanical data, i.e., lattice parameters, aver-age cohesive energy, average bond length, average height, 2D elastic constants as well as phonon data are determined here using a single consistent first-principles approach as described in the next

Section 2.1. These data were further considered as reference data and marked as Value$^{DFT}$. Then the same data were determined, as described in Section 2.2 using the analyzed molecular potentials from Subsection 2.2.1 and are marked as Value$^{potential}$. Having both data, we can simply define mean absolute percentage error (MAPE):

$$\text{MAPE} = \frac{100\%}{n} \sum_{t=1}^{n} \frac{\left| \text{Value}^{DFT} - \text{Value}^{potential} \right|}{\text{Value}^{DFT}}, \tag{1}$$

that will allow us to quantify the potentials under examination.

For 2D materials, directional 2D Young's moduli

$$E^{2D}_{[10]} = \frac{C_{11}C_{22} - C_{12}^2}{C_{22}} \quad \text{and} \quad E^{2D}_{[01]} = \frac{C_{11}C_{22} - C_{12}^2}{C_{11}}, \tag{2}$$

2D Poisson's ratios

$$\nu^{2D}_{[10]} = \frac{C_{12}}{C_{22}} \quad \text{and} \quad \nu^{2D}_{[01]} = \frac{C_{12}}{C_{11}}, \tag{3}$$

and the 2D shear modulus

$$G^{2D} = C_{33}, \tag{4}$$

are often used instead of elastic constants $C_{ij}$. Due to symmetry for hexagonal lattices, this reduces to one 2D Young's modulus E and a one 2D Poisson's ratio , see [13].

## 2.1 *Ab Initio* Calculations

The *ab initio* calculation methodology here is closely analogous to that used in [14], i.e., the den-sity functional theory (DFT) [15,16], ABINIT plane-wave approximation code [17,18], local den-sity approximation (LDA) [19,20] as an exchange-correlation (XC) functional, optimized norm-conserving Vanderbilt pseudopotential [21] (ONCVPP). The *cut-off* energy and the electron con-figuration for Si used in the DFT calculations according to the pseudopotential and Gaussian smearing scheme with *tsmear* (Ha)=0.02 were used. To generate K-PoinTs grids *kptrlen* was set to 43.0. Since 2D structures were analyzed in the z-direction, a vacuum of 20 Å was applied. The initial data for the five structures analyzed were deduced from [6] and [7]. The structures were then carefully relaxed with full optimization of cell geometry and atomic coordinates [14].

The average cohesive energy $E_c$ (eV/atom) was computed as the difference in the total energy of a given relaxed earlier structure and its individual atoms placed in a cubic box of sufficient size. The theoretical ground state, T=0 K, elastic constants, $C_{ij}$, of all the previously optimized structures were computed using the metric tensor formulation of strain in density functional perturbation the-ory (DFPT) [22]. The mechanical stability of the analyzed structures was verified by calculating the so-called Kelvin moduli [23-25]. To calculate the phonons, the density functional perturbation theory implemented in ABINIT [17,18] was employed. The phonon dispersion curves along the path Γ[0,0,0]-**M**[1/2,0,0]-**K**[1/3,1/3,0]-Γ[0,0,0] [26] of the analyzed structures were then used to identify their dynamical stability [27], complementary to the mechanical stability.

## 2.2 Molecular Calculations

To perform molecular calculations the molecular statics (MS) method, T=0 K, [28-30] was used by means of the Large-scale Atomic/Molecular Massively Parallel Simulator (LAMMPS) [31] and analyzed by means of the Open Visualization Tool (OVITO) [32]. As for DFT calculations, the structures were here fully pre–relaxed with the conjugate gradient (CG) algorithm and then the elastic constants, $C_{ij}$, were calculated for them using the stress–strain method with the maximum strain magnitude set to $10^{-6}$ [30,31]. In the z-direction, a vacuum was set to 20 Å.

To measure performance of the interatomic potentials analyzed, the series of molecular dynamics (MD) simulations (200 atoms and 10000 timesteps, NVT ensemble) and LAMMPS's built-in func-tion *timesteps/s* were used. The results were then normalized relative to the longest run time.

### 2.2.1 Interatomic Potentials

The parameterizations of the potentials listed below were obtained from the NIST Interatomic Potentials Repository and/or from LAMMPS code sources.

1. **Tersoff1988** [33]: the original Tersoff potential for silicon (it is important to remember that this paper proposed a form of potential rather than a specific parametrization for silicon)

2. **Tersoff2007** [34]: the Tersoff potential fitted to the elastic constants of diamond silicon

3. **Tersoff2017** [35]: newer, better optimized the Tersoff potential for silicon

4. **MEAM2007** [36]: a semi-empirical interatomic potential for silicon based on the modified embedded atom method (MEAM) formalism

5. **MEAM2011** [37]: spline-based modified embedded-atom method (MEAM) potential for Si fitted for silicon interstitials

6. **SW1985** [38]: the Stillinger–Weber (SW) potential fitted to solid and liquid forms of Si

7. **SW2014** [39]: the Stillinger–Weber (SW) potential fitted to phonon dispersion curves of a single-layer Si sheet

8. **EDIP** [40]: the environment-dependent interatomic potential (EDIP) fitted to various bulk phases and defect structures of Si

9. **ReaxFF** [41]: the reactive force-field (ReaxFF) fitted to a training set of DFT data that per-tain to Si/Ge/H bonding environments

10. **COMB** [42]: the charge optimized many-body (COMB) potential fitted to a pure silicon and five polymorphs of silicon dioxide

11. **SNAP** [43]: the machine-learning-based (ML-IAP) linear variant of spectral neighbor analy-sis potential (SNAP) fitted to total energies and interatomic forces in ground-state Si, strained structures and slab structures obtained from DFT calculations

12. **qSNAP** [43]: the machine-learning-based (ML-IAP) quadratic variant of spectral neighbor analysis potential (qSNAP) fitted to total energies and interatomic forces in ground-state Si, strained structures and slab structures obtained from DFT calculations

13. **SO(3)** [44]: the machine-learning-based (ML-IAP) variant of the SO(3) smooth power spec-trum potential (SO(3)) fitted to the ground-state of the crystalline silicon structure, strained structures, slabs, vacancy, and liquid configurations from DFT simulations

14. **ACE** [45]: the machine-learning-based (ML-IAP) variant of the atomic cluster expansion potential (ACE) fitted to a wide range of properties of 3D silicon determined from the DFT calculation

# 3 Results and Discussion

## 3.1 Structural and Mechanical Properties

Basic cells for the five silicene polymorphs, i.e, the flat (FS):(*hP2*, P6/mmm, no.191), low-buckled (LBS):(*hP2*, P-3m1, no.164), trigonal dumbbell (TDS):(*hP7*, P-62m, no.189), honeycomb dumb-bell (HDS):(*hP8*, P6/mmm, no.191) and large honeycomb dumbbell (LHDS):(*hP10*, P6/mmm, no.191) are depicted in Figure 1 and additionally the crystallographic data for them are stored in Crystallographic Information Files (CIFs) in the Supporting Information.

The results of first-principles calculations show that all the silicene phases have hexagonal sym-metry. The symmetry characteristic of a structure determines the symmetry of its physical proper-ties (*Neumann's Principle* and *Curie laws*) [46,47]. For 2D linear hyperelastic materials, there are four classes of symmetry [24] and hexagonal symmetry implies isotropy of the stiffness tensor, i.e., there are only two distinct elastic constants and they satisfy, in Voigt notation, such conditions that $C_{11} = C_{22}$, $C_{33} = (C_{11} − C_{12})/2$.

Determined from DFT computations structural and mechanical characteristics, namely lattice pa-rameters, average cohesive energy, average bond length, average height, 2D elastic constants, 2D Young's modulus, Poisson's ratio and 2D Kelvin moduli, of the five silicene polymorphs analyzed are gathered in Table 1. Since we are analyzing free-standing silicene here, which has not yet been observed in experiments, we compare the results of the calculations with those of other authors. We find that the lattice constants, average bond length, average height and cohesive energy agree at the DFT level of accuracy with other calculations. Mechanical properties of silicene are available in the literature only for the LBS phase and are limited to 2D Young's modulus and Poisson's ratio only. These quantities are also in reasonable agreement with the present results. It is immediately worth noting that all the calculated 2D Kelvin moduli for all silicene phases are positive, which results in mechanical stability [24]. The phonon spectra along the Γ-**M**-**K**-Γ path for the five silicene polymorphs is depicted in Fig-ure 2. The analysis of the computed curves shows that the phases TDS,LBS,HDS and LHDS are not only mechanically but also dynamically stable, i.e., all phonon modes have positive frequencies anywhere. The FS phase is mechanically stable, but can be dynamically unstable, i.e., optical ZO phonon mode has a negative frequency. Other authors have also observed similar FS phase behav-ior [4,12], however, but since silicene is not a free-standing structure in nature the selection of a proper substrate may dampen the out-of-plane vibration mode and a flat silicene may be produced [3]. So it was decided that this phase was also included in the molecular calculations.

## 3.2 Performance of Interatomic Potentials

The computed with the use of molecular statics and the fourteen various interatomic potentials for silicon (**Tersoff** (×3), **MEAM** (×2), **Stillinger-Weber** (×2), **EDIP**, **ReaxFF**, **COMB** and machine-learning-based (**ML-IAP** (×4)), enumerated in Section 2.2.1, twelve structural and mechanical properties, i.e., lattice parameters a, b, average cohesive energy , average bond length , average height $h$, 2D elastic constants $C_{ij}$, 2D Kelvin moduli of the flat silicene (FS) phase are collected in Table 2, of the low-buckled silicene (LBS) in Table 3, of the trigonal dumbbell silicene (TDS) in Table 4, of the honeycomb dumbbell silicene (HDS) in Table 5 and of the large honeycomb dumb-bell silicene (LHDS) in Table 6, respectively. The aforementioned results, for each of the five sil-icene phases, were then compared with those from the DFT calculations using the mean absolute percentage error (MAPE) defined in Equation 1. Let's briefly analyze the results for each phase, and so for the FS phase the most accurate is the **MEAM2011** potential, it has the lowest MAPE, see Table 2, for the LBS phase the **Tersoff2107**, see Table 3, for the TDS phase the **ReaxFF**, see Table 4, for the HDS the **ReaxFF**, see Table 5 and for the LHDS again the **ReaxFF** potential, see Table 6, respectively. Now let's take a summary look. Seven of the fourteen potentials analyzed, namely **Tersoff2007**, **Tersoff2017**, **SW1985**, **SW2014**, **ReaxFF**, **SNAP** and **ACE**, are able to cor-rectly reproduce the structural properties of the five polymorphs of silicene, see Tabs. 3, 5, 6. Two potentials: **ReaxFF** and **MEAM2011** give the best quantitative performance measured by the total mean absolute percentage error (MAPE), see Tab. 6. From the point of view of the cost of calculations in terms of relative performance measured as normalized timesteps/second in molecular dynamics (MD) simulation the **EDIP** and **Stillinger-Weber** potentials are the fastest, about 5 times faster than the **MEAM** and **Tersoff** potentials, about 100 times faster than **ReaxFF** and **COMB**, and up to 2000 times faster than the **ML-IAP(ACE)** potential, see Tab. 6. It is also worth noting, that the machine-learning-based (**ML-IAP**) interatomic potentials, according to the methodology used, are not superior to classical potentials in terms of performance (MAPE) and are instead even three orders of magnitude more computationally expensive, see Tab. 6.

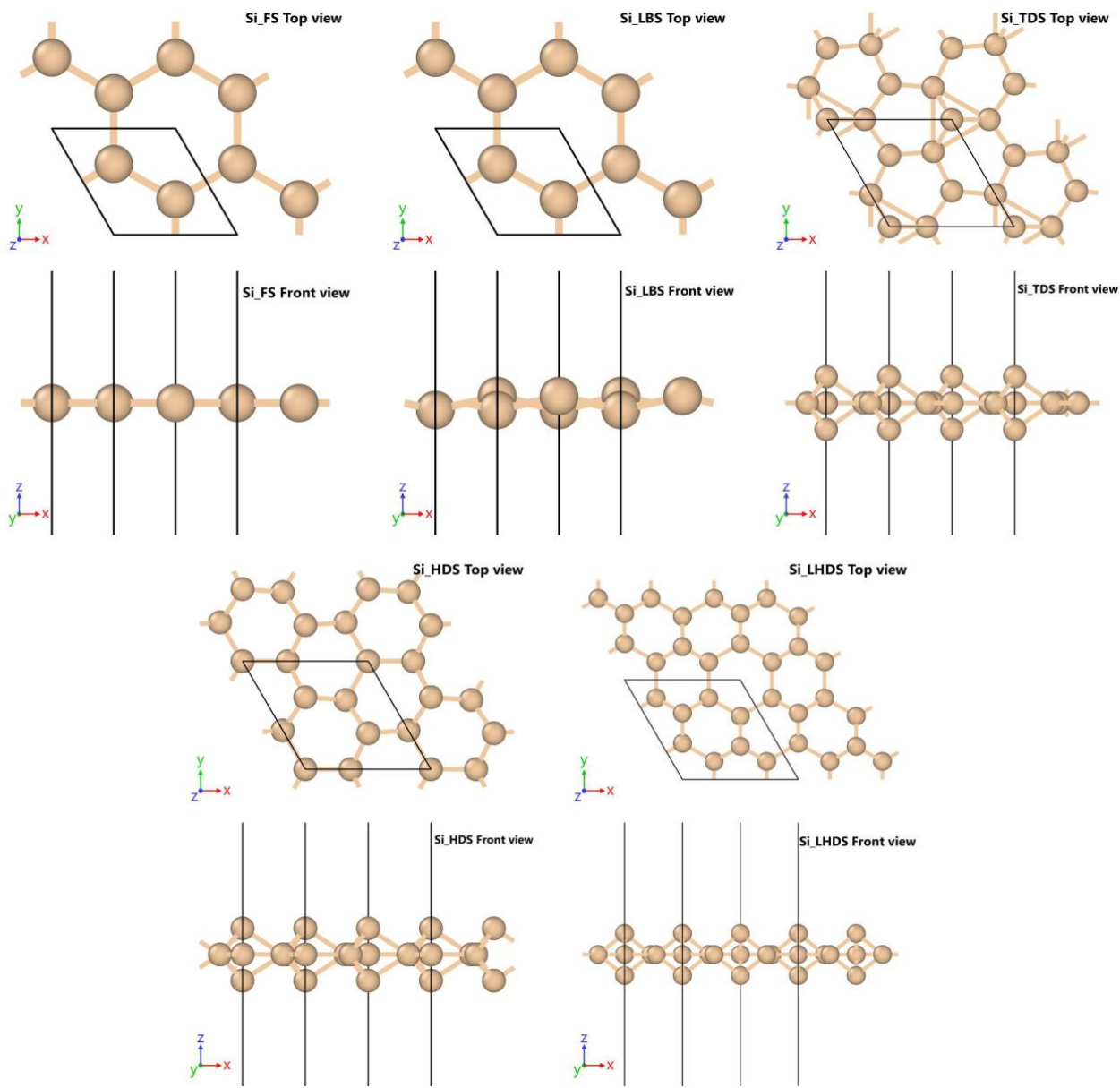

**Figure 1:** Polymorphs of silicene.

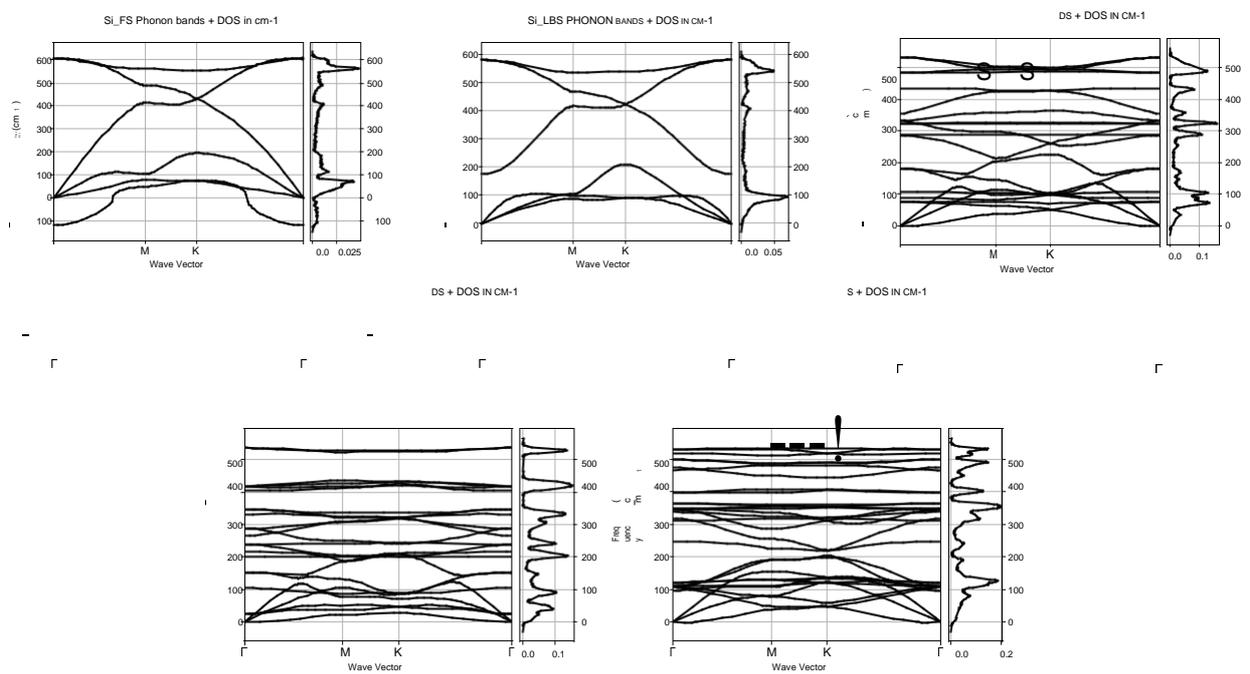

**Figure 2:** Phonon dispersion and density of states (DOS) of the flat (FS), low-buckled (LBS), trigonal dumbbell (TDS), honeycomb dumbbell (HDS) and large honeycomb dumbbell (LHDS) single-layer silicon (silicene) phases. High symmetry points: Γ[0,0,0], **M**[1/2,0,0], **K**[1/3,1/3,0].

Table 1: Structural and mechanical properties of flat (FS), low-buckled (LBS), trigonal dumbbell (TDS), honeycomb dumbbell (HDS) and large honeycomb dumbbell (LHDS) silicene phases from density functional theory (DFT) calculations: lattice parameters a,b (Å), average cohesive energy (eV/atom), average bond length (Å), average height $h$ (Å), 2D elastic constants $C_{ij}$ (N/m), 2D Young's modulus E (N/m), Poisson's ratio and 2D Kelvin moduli (N/m).

| Polymorph | FS | | LBS | | TDS | | HDS | | LHDS | |
|---|---|---|---|---|---|---|---|---|---|---|
| Source | This work | Refs. | This work | Refs. | This work | Refs. | This work | Refs. | This work | Refs. |
| a | 3.855 | 3.90[a] | 3.828 | 3.87[a], 3.83[b] | 6.434 | 6.52[b] | 6.297 | 6.38[b] | 7.334 | 7.425[b] |
| b | 3.855 | 3.90[a] | 3.828 | 3.87[a], 3.83[b] | 6.434 | 6.52[b] | 6.297 | 6.38[b] | 7.334 | 7.425[b] |
| — | 4.562 | 4.764[a] | 4.577 | 4.784[a], 5.16[b] | 4.679 | | 4.679 | | 4.769 | |
| † | 2.225 | | 2.249 | 2.25[b] | 2.331 | | 2.399 | | 2.357 | |
| $h$ | 0.0 | 0.0[a] | 0.421[a] | 0.45[a], 0.44[b] | 2.734 | | 2.635 | | 2.683 | |
| $C_{11}$ | 84.8 | | 69.2 | | 100.5 | | 141.6 | | 104.5 | |
| $C_{22}$ | 84.8 | | 69.2 | | 100.5 | | 141.6 | | 104.5 | |
| $C_{12}$ | 40.6 | | 22.1 | | 52.3 | | 96.4 | | 52.7 | |
| $C_{33}$ | 22.1 | | 23.6 | | 24.1 | | 22.6 | | 25.9 | |
| E | 65.4 | | 62.2 | 61.8[a] | 73.3 | | 76.0 | | 77.9 | |
| | 0.48 | | 0.32 | 0.31[a] | 0.52 | | 0.68 | | 0.50 | |
| | 125.4 | | 91.3 | | 152.8 | | 238.0 | | 157.2 | |
| | 44.3 | | 47.1 | | 48.3 | | 45.2 | | 51.8 | |
| | 44.3 | | 47.1 | | 48.3 | | 45.2 | | 51.8 | |

[a] Ref. [48], [b] Ref. [7].
† An average bond lengths calculated using radial pair distribution function with a *cut-off* radius = 3.0 Å and a number of histogram bins = 1000 [32].

**Table 2:** Structural and mechanical properties of flat silicene (FS) from molecular calculations: lattice parameters a, b (Å), average cohesive energy (eV/atom), average bond length (Å), average height $h$ (Å), 2D elastic constants $C_{ij}$ (N/m), 2D Kelvin moduli (N/m), mean absolute percentage error (MAPE) (%).

| Method | DFT | Tersoff 1988 | Tersoff 2007 | Tersoff 2017 | MEAM 2007 | MEAM 2011 | SW 1985 | SW 2014 | EDIP | ReaxFF | COMB | ML-IAP SNAP | ML-IAP qSNAP | ML-IAP SO(3) | ML-IAP ACE |
|---|---|---|---|---|---|---|---|---|---|---|---|---|---|---|---|
| a | 3.855 | 4.008 | 4.019 | 4.042 | 4.457 | 3.960 | 4.104 | 3.886 | 4.018 | 3.950 | 3.990 | 4.121 | 4.019 | 4.051 | 3.850 |
| b | 3.855 | 4.008 | 4.019 | 4.042 | 4.457 | 3.960 | 4.104 | 3.886 | 4.018 | 3.950 | 3.990 | 4.121 | 4.019 | 4.051 | 3.850 |
| $\bar{E}$ | 4.562 | 3.926 | 3.828 | 3.687 | 3.288 | 3.793 | 3.145 | 2.564 | 4.010 | 3.408 | 3.911 | 4.575 | 4.499 | 4.407 | 0.962 |
| $\bar{d}$ | 2.225 | 2.315 | 2.321 | 2.333 | 2.573 | 2.288 | 2.369 | 2.243 | 2.321 | 2.282 | 2.306 | 2.381 | 2.321 | 2.340 | 2.222 |
| $h$ | 0.0 | 0.0 | 0.0 | 0.0 | 0.0 | 0.0 | 0.0 | 0.0 | 0.0 | 0.0 | 0.0 | 0.0 | 0.0 | 0.0 | 0.0 |
| $C_{11}$ | 84.8 | 54.6 | 55.2 | 47.7 | 57.3 | 84.0 | 58.8 | 57.3 | 87.4 | 74.7 | 79.7 | 41.5 | 24.7 | 41.0 | 88.9 |
| $C_{22}$ | 84.8 | 54.6 | 55.2 | 47.7 | 57.3 | 84.0 | 58.8 | 57.3 | 87.4 | 74.7 | 79.7 | 41.5 | 24.7 | 41.0 | 88.9 |
| $C_{12}$ | 40.6 | 49.5 | 47.4 | 52.8 | 29.7 | 40.8 | 34.4 | 33.0 | 9.2 | 36.8 | 23.9 | 16.2 | 17.6 | 23.9 | 44.8 |
| $C_{33}$ | 22.1 | 2.6 | 3.9 | -2.6 | 13.8 | 21.6 | 12.2 | 12.2 | 39.1 | 19.0 | 27.9 | 12.6 | 3.5 | 8.6 | 22.0 |
| $\lambda_1$ | 125.4 | 104.0 | 102.6 | 100.5 | 87.0 | 124.8 | 93.2 | 90.3 | 96.6 | 111.5 | 103.6 | 57.6 | 42.3 | 64.9 | 133.7 |
| $\lambda_2$ | 44.3 | 5.1 | 7.8 | -5.1 | 27.6 | 43.3 | 24.4 | 24.4 | 78.2 | 37.9 | 55.8 | 25.3 | 7.0 | 17.1 | 44.1 |
| $\lambda_3$ | 44.3 | 5.1 | 7.8 | -5.1 | 27.6 | 43.3 | 24.4 | 24.4 | 78.2 | 37.9 | 55.8 | 25.3 | 7.0 | 17.1 | 44.1 |
| MAPE$_{FS}$ |  | 36.515 | 34.619 | 45.987 | 28.174 | 3.147 | 26.093 | 26.595 | 32.827 | 10.904 | 15.805 | 33.284 | 48.283 | 35.946 | 9.757 |

**Table 3:** Structural and mechanical properties of low-buckled silicene (LBS) from molecular calculations: lattice parameters a, b (Å), average cohesive energy (eV/atom), average bond length (Å), average height $h$ (Å), 2D elastic constants $C_{ij}$ (N/m), 2D Kelvin moduli (N/m), mean absolute percentage error (MAPE) (%).

| Method | DFT | Tersoff 1988 | Tersoff 2007 | Tersoff 2017 | MEAM 2007 | MEAM 2011 | SW 1985 | SW 2014 | EDIP | ReaxFF | COMB | ML-IAP SNAP | ML-IAP qSNAP | ML-IAP SO(3) | ML-IAP ACE |
|---|---|---|---|---|---|---|---|---|---|---|---|---|---|---|---|
| a | 3.828 | 3.309 | 3.820 | 3.870 | 4.150 | 3.837 | 3.840 | 3.812 | 4.018 | 3.843 | 3.990 | 3.916 | 3.808 | 3.743 | 3.702 |
| b | 3.828 | 3.309 | 3.820 | 3.870 | 4.150 | 3.837 | 3.840 | 3.812 | 4.018 | 3.843 | 3.990 | 3.916 | 3.808 | 3.743 | 3.702 |
| $\bar{E}$ | 4.577 | 3.936 | 3.936 | 3.755 | 3.404 | 3.851 | 3.252 | 2.572 | 4.010 | 3.454 | 3.911 | 4.648 | 4.624 | 4.606 | 0.926 |
| $\bar{d}$ | 2.249 | 2.315 | 2.312 | 2.315 | 2.534 | 2.297 | 2.351 | 2.243 | 2.321 | 2.300 | 2.306 | 2.390 | 2.366 | 2.347 | 2.242 |
| $h$ | 0.421 | 1.304 | 0.690 | 0.327 | 0.820 | 0.608 | 0.784 | 0.427 | 0.0$^\ddagger$ | 0.610 | 0.0$^\ddagger$ | 0.774 | 0.859 | 0.913 | 0.675 |
| $C_{11}$ | 69.2 | 0.006 | 59.6 | 50.1 | 38.4 | 47.1 | 36.4 | 30.8 | 87.4 | 49.0 | 79.7 | 24.8 | 28.8 | 31.7 | 55.0 |
| $C_{22}$ | 69.2 | 0.006 | 59.6 | 50.1 | 38.4 | 47.1 | 36.4 | 30.8 | 87.4 | 49.0 | 79.7 | 24.8 | 28.8 | 31.7 | 55.0 |
| $C_{12}$ | 22.1 | 0.002 | 4.1 | 5.5 | 4.4 | 10.6 | 6.1 | 5.4 | 9.2 | 14.1 | 23.9 | 5.6 | 6.3 | 9.0 | 2.7 |
| $C_{33}$ | 23.6 | 0.002 | 27.7 | 22.3 | 17.0 | 18.3 | 15.1 | 12.7 | 39.1 | 17.5 | 27.9 | 9.6 | 11.2 | 11.4 | 26.1 |
| $\lambda_1$ | 91.3 | 0.008 | 63.7 | 55.7 | 42.7 | 57.7 | 42.6 | 36.2 | 96.6 | 63.1 | 103.6 | 30.4 | 35.1 | 40.6 | 57.7 |
| $\lambda_2$ | 47.1 | 0.004 | 55.5 | 44.6 | 34.0 | 36.5 | 30.3 | 25.5 | 78.2 | 34.9 | 55.8 | 19.2 | 22.4 | 22.7 | 52.3 |
| $\lambda_3$ | 47.1 | 0.004 | 55.5 | 44.6 | 34.0 | 36.5 | 30.3 | 25.5 | 78.2 | 34.9 | 55.8 | 19.2 | 22.4 | 22.7 | 52.3 |
| MAPE$_{LBS}$ |  | 79.467 | 22.781 | 19.235 | 37.991 | 23.569 | 37.306 | 35.911 | 36.630 | 22.977 | 19.433 | 45.306 | 43.176 | 42.060 | 28.792 |

$^\ddagger$ Input LBS converges to FS.

**Table 4:** Structural and mechanical properties of trigonal dumbbell silicene (TDS) from molecular calculations: lattice parameters a, b (Å), average cohesive energy (eV/atom), average bond length (Å), average height $h$ (Å), 2D elastic constants $C_{ij}$ (N/m), 2D Kelvin moduli (N/m), mean absolute percentage error (MAPE) (%).

| Method | DFT | Tersoff 1988 | Tersoff 2007 | Tersoff 2017 | MEAM 2007 | MEAM 2011 | SW 1985 | SW 2014 | EDIP | ReaxFF | COMB | ML-IAP SNAP | ML-IAP qSNAP | ML-IAP SO(3) | ML-IAP ACE |
|---|---|---|---|---|---|---|---|---|---|---|---|---|---|---|---|
| a | 6.434 | 6.480 | 6.471 | 6.475 | 7.140 | 6.511 | 6.600 | 6.291 | 6.759 | 6.380 | 6.574 | 6.777 | 6.797 | 6.518 | 6.448 |
| b | 6.434 | 6.480 | 6.471 | 6.475 | 7.140 | 6.511 | 6.600 | 6.291 | 6.759 | 6.380 | 6.574 | 6.777 | 6.797 | 6.518 | 6.448 |
| $\bar{E}$ | 4.679 | 4.248 | 3.865 | 3.890 | 3.427 | 3.865 | 3.322 | 2.591 | 4.075 | 3.551 | 3.723 | 4.726 | 4.693 | 4.607 | 0.4606 |
| $\bar{d}$ | 2.331 | 2.362 | 2.371 | 2.362 | 2.590 | 2.376 | 2.431 | 2.315 | 2.416 | 2.344 | 2.387 | 2.398 | 2.404 | 2.505 | 2.429 |
| $h$ | 2.734 | 2.870 | 3.110 | 3.056 | 3.228 | 2.856 | 3.260 | 3.100 | 2.628 | 3.115 | 2.994 | 2.518 | 2.540 | 3.058 | 2.649 |
| $C_{11}$ | 100.5 | 79.4 | 67.9 | 51.4 | 69.9 | 80.5 | 78.7 | 68.4 | 60.3 | 94.6 | 70.8 | 44.9 | 36.7 | 35.3 | 90.5 |
| $C_{22}$ | 100.5 | 79.4 | 67.9 | 51.4 | 69.9 | 80.5 | 78.7 | 68.4 | 60.3 | 94.6 | 70.8 | 44.9 | 36.7 | 35.3 | 90.5 |
| $C_{12}$ | 52.3 | 63.2 | 39.4 | 39.0 | 34.9 | 31.0 | 39.6 | 35.1 | 30.2 | 39.7 | 25.6 | 16.1 | 16.7 | 10.2 | 33.0 |
| $C_{33}$ | 24.1 | 8.1 | 14.3 | 6.2 | 17.5 | 24.8 | 19.6 | 16.7 | 15.1 | 27.4 | 22.6 | 14.4 | 10.0 | 12.5 | 28.8 |
| $\lambda_1$ | 152.8 | 142.7 | 107.3 | 90.5 | 104.8 | 111.5 | 118.2 | 103.5 | 90.5 | 134.3 | 96.4 | 61.1 | 53.4 | 45.5 | 123.5 |
| $\lambda_2$ | 48.3 | 16.2 | 28.5 | 12.4 | 35.0 | 49.5 | 39.1 | 33.3 | 30.1 | 54.9 | 45.2 | 28.8 | 20.1 | 25.0 | 57.6 |
| $\lambda_3$ | 48.3 | 16.2 | 28.5 | 12.4 | 35.0 | 49.5 | 39.1 | 33.3 | 30.1 | 54.9 | 45.2 | 28.8 | 20.1 | 25.0 | 57.6 |
| MAPE$_{TDS}$ |  | 23.790 | 22.993 | 34.815 | 23.821 | 11.809 | 17.077 | 23.723 | 25.537 | 10.755 | 16.881 | 31.929 | 38.100 | 37.357 | 19.317 |

**Table 5:** Structural and mechanical properties of honeycomb dumbbell silicene (HDS) from molecular calculations: lattice parameters a, b (Å), average cohesive energy (eV/atom), average bond length (Å), average height $h$ (Å), 2D elastic constants $C_{ij}$ (N/m), 2D Kelvin moduli (N/m), mean absolute percentage error (MAPE) (%).

| Method | DFT | Tersoff 1988 | Tersoff 2007 | Tersoff 2017 | MEAM 2007 | MEAM 2011 | SW 1985 | SW 2014 | EDIP | ReaxFF | COMB | ML-IAP SNAP | ML-IAP qSNAP | ML-IAP SO(3) | ML-IAP ACE |
|---|---|---|---|---|---|---|---|---|---|---|---|---|---|---|---|
| a | 6.297 | 6.074† | 6.133 | 6.114 | 5.608† | 6.364† | 6.272 | 6.063 | 6.349† | 6.048 | 6.344 | 6.644 | 6.510† | 5.030† | 6.250 |
| b | 6.297 | 5.864† | 6.133 | 6.114 | 6.127† | 6.279† | 6.272 | 6.063 | 6.389† | 6.048 | 6.344 | 6.644 | 6.344† | 5.818† | 6.250 |
| – | 4.679 | 4.334 | 3.738 | 3.816 | 3.663 | 4.008 | 3.243 | 2.467 | 4.105 | 3.366 | 3.481 | 4.766 | 4.699 | 4.873 | 0.391 |
| – | 2.399 | 2.398 | 2.425 | 2.408 | 2.641 | 2.405 | 2.495 | 2.394 | 2.489 | 2.396 | 2.511 | 2.436 | 2.457 | 2.612 | 2.371 |
| $h$ | 2.635 | 2.596 | 3.050 | 3.006 | 3.500 | 2.990 | 3.189 | 3.010 | 2.365 | 2.952 | 2.989 | 2.404 | 2.438 | 2.162 | 2.559 |
| $C_{11}$ | 141.6 | 89.3 | 59.4 | 64.0 | 56.2 | 78.2 | 57.4 | 35.6 | 78.0 | 75.4 | 14.3 | 41.9 | 30.8 | 76.5 | 69.6 |
| $C_{22}$ | 141.6 | 46.3 | 59.4 | 64.0 | 57.1 | 82.4 | 57.4 | 35.6 | 45.5 | 75.4 | 14.3 | 41.9 | 16.0 | 73.0 | 69.6 |
| $C_{12}$ | 96.4 | 20.4 | 24.6 | 31.3 | 17.3 | 7.1 | 25.3 | 14.9 | 46.3 | 31.4 | -37.4 | 16.2 | 13.8 | -10.2 | 10.6 |
| $C_{33}$ | 22.6 | 24.8 | 17.4 | 16.3 | 16.7 | 26.2 | 16.0 | 10.4 | 9.6 | 22.0 | 25.8 | 12.9 | 11.7 | 2.7 | 29.5 |
|  | 238.0 | 102.1 | 84.0 | 95.3 | 76.6 | 87.8 | 82.7 | 50.6 | 119.7 | 106.8 | 51.7 | 58.1 | 41.2 | 84.9 | 80.1 |
|  | 45.2 | 57.7 | 34.8 | 32.7 | 44.5 | 72.9 | 32.0 | 20.7 | 13.6 | 43.9 | -23.1★ | 25.7 | 24.3 | 64.7 | 59.0 |
|  | 45.2 | 25.5 | 34.8 | 32.7 | -1.2★ | 52.3 | 32.0 | 20.7 | 9.2 | 43.9 | 51.7 | 25.7 | 23.4 | 5.5 | 59.0 |
| MAPE$_{HDS}$ |  | 28.361 | 30.540 | 29.931 | 39.880 | 30.392 | 33.503 | 45.400 | 37.500 | 22.740 | 51.780 | 37.684 | 41.062 | 45.571 | 37.134 |

† Potential does not reproduce the correct symmetry of the structure (a≠b),
★ Negative Kelvin moduli indicating a lack of mechanical stability.

**Table 6:** Structural and mechanical properties of large honeycomb dumbbell silicene (LHDS) from molecular calculations: lattice parameters a, b (Å), average cohesive energy (eV/atom), average bond length (Å), average height $h$ (Å), 2D elastic constants $C_{ij}$ (N/m), 2D Kelvin moduli (N/m), mean absolute percentage error (MAPE) (%), relative performance measured as normalized timesteps/second in molecular dynamics (MD) simulation.

| Method | DFT | Tersoff 1988 | Tersoff 2007 | Tersoff 2017 | MEAM 2007 | MEAM 2011 | SW 1985 | SW 2014 | EDIP | ReaxFF | COMB | ML-IAP SNAP | ML-IAP qSNAP | ML-IAP SO(3) | ML-IAP ACE |
|---|---|---|---|---|---|---|---|---|---|---|---|---|---|---|---|
| a | 7.334 | 7.000† | 7.249 | 7.236 | 7.900† | 7.363 | 7.403 | 7.062 | 7.705 | 7.167 | 7.422 | 7.648 | 7.741 | 7.427 | 7.393 |
| b | 7.334 | 6.978† | 7.249 | 7.236 | 7.560† | 7.363 | 7.403 | 7.062 | 7.705 | 7.167 | 7.422 | 7.648 | 7.741 | 7.427 | 7.393 |
| – | 4.769 | 4.468 | 3.897 | 4.004 | 3.505 | 3.911 | 3.399 | 2.602 | 4.113 | 3.623 | 3.646 | 4.804 | 4.794 | 4.698 | 0.370 |
| – | 2.357 | 2.381 | 2.387 | 2.369 | 2.627 | 2.407 | 2.456 | 2.345 | 2.438 | 2.370 | 2.454 | 2.436 | 2.403 | 2.385 | 2.515 |
| $h$ | 2.683 | 2.692 | 3.109 | 3.050 | 3.050 | 2.857 | 3.250 | 3.100 | 2.637 | 3.112 | 2.994 | 2.538 | 2.562 | 2.700 | 2.712 |
| $C_{11}$ | 104.5 | 2.3 | 78.5 | 68.6 | 45.6 | 88.0 | 84.5 | 73.0 | 53.7 | 99.0 | 53.1 | 44.6 | 43.8 | 25.9 | 59.8 |
| $C_{22}$ | 104.5 | 7.3 | 78.5 | 68.6 | 56.2 | 88.0 | 84.5 | 73.0 | 53.7 | 99.0 | 53.1 | 44.6 | 43.8 | 25.9 | 59.8 |
| $C_{12}$ | 52.7 | -13.1 | 42.3 | 25.5 | 22.0 | 29.9 | 46.1 | 38.9 | 36.0 | 40.4 | 31.7 | 16.1 | 19.1 | 0.4 | 15.2 |
| $C_{33}$ | 25.9 | 9.0 | 18.1 | 21.6 | 14.0 | 29.0 | 19.2 | 17.1 | 8.8 | 29.3 | 10.7 | 14.2 | 12.3 | 12.7 | 22.3 |
|  | 157.2 | 18.5 | 120.8 | 94.1 | 73.6 | 117.9 | 130.6 | 111.9 | 89.7 | 139.4 | 84.8 | 60.6 | 62.8 | 26.4 | 75.0 |
|  | 51.8 | 17.8 | 36.2 | 43.1 | 28.8 | 58.1 | 38.4 | 34.1 | 17.6 | 58.6 | 21.5 | 28.5 | 24.7 | 25.5 | 44.7 |
|  | 51.8 | -8.5★ | 36.2 | 43.1 | 27.4 | 58.1 | 38.4 | 34.1 | 17.6 | 58.6 | 21.5 | 28.5 | 24.7 | 25.5 | 44.7 |
| MAPE$_{LHDS}$ |  | 55.693 | 18.373 | 20.287 | 34.499 | 13.636 | 16.741 | 23.855 | 33.228 | 10.809 | 33.466 | 33.219 | 34.603 | 40.928 | 29.310 |
| $\sum$MAPE |  | 223.826 | 129.306 | 150.256 | 164.364 | 82.553 | 130.721 | 155.483 | 165.722 | 78.185 | 137.365 | 181.422 | 205.224 | 201.861 | 124.310 |
| timesteps/s |  | 387.2 | 382.7 | 355.2 | 505.7 | 416.9 | 904.9 | 1753.8 | 2032.1 | 23.6 | 26.4 | 7.9 | 4.7 | 4.2 | 1.0 |

† Potential does not reproduce the correct symmetry of the structure (a≠b),
★ Negative Kelvin moduli indicating a lack of mechanical stability.

# 4  Conclusion

A systematic quantitative comparative study of various silicon interatomic potentials for repro-ducing the properties of five silicene (2D silicon) polymorphs was shown. In order to compare the fourteen potentials listed in Section 2.2.1, the structural and mechanical properties of flat (FS), low-buckled (LBS), trigonal dumbbell (TDS), honeycomb dumbbell (HDS) and large honeycomb dumbbell (LHDS) silicene (Fig. 1) obtained from the density functional theory (DFT) and molecu-lar statics (MS) computations were used. The computational cost, the performance, of the analyzed potentials were also compared.

It can be stated that considering the performance and the cost of calculations, the classical poten-tials of Tersoff, SW and MEAM type seem to be the best choice here. Although data for silicene polytypes were not used in the optimization of these potentials, they were able to reproduce their properties well. There is a consequence that they are based on physics, have a natural extrapolation ability, and not just interpolate data.

I hope that the findings done here will help other researchers in selecting the suitable potentials for their purposes and will be a hint to parameterize new potentials for silicene.

# Supporting Information

Crystallographic Information Files (CIF) for polymorphs of silicene.

Supporting Information File 1:

File Name: Si_FS.cif

File Format: CIF

Title: flat silicene (FS)

Supporting Information File 2:

File Name: Si_LBS.cif

File Format: CIF

Title: low-buckled silicene (LBS)

Supporting Information File 3:

File Name: Si_TDS.cif

File Format: CIF

Title: trigonal dumbbell silicene (TDS)

Supporting Information File 4:

File Name: Si_HDS.cif

File Format: CIF

Title: honeycomb dumbbell silicene (HDS)

Supporting Information File 5:

File Name: Si_LHDS.cif

File Format: CIF

Title: large honeycomb dumbbell silicene (LHDS)

# Si_FS.cif -- in P1.1 -- non-magnetic

data_Si_FS_191

_audit_creation_date                222-8-4

```
_audit_creation_method              "Bilbao Crystallographic Server"
_symmetry_Int_Tables_number         1
#_symmetry_space_group_name_H-M     "P1'"
_cell_length_a                      3.85481
_cell_length_b                      3.85481
_cell_length_c                      2.
_cell_angle_alpha                   9.
_cell_angle_beta                    9.
_cell_angle_gamma                   12.

loop_
_symmetry_equiv_pos_site_id
_symmetry_equiv_pos_as_xyz
   1    x,y,z

loop_
_atom_site_label
_atom_site_type_symbol
_atom_site_fract_x
_atom_site_fract_y
_atom_site_fract_z
_atom_site_occupancy
Si1 Si .33333 .66667 .5 1.
Si2 Si .66666 .33333 .5 1.

#  Si_LBS.cif -- in P1.1 -- non-magnetic

data_Si_LBS_164
```

```
_audit_creation_date                  222-8-4
_audit_creation_method                "Bilbao Crystallographic Server"
_symmetry_Int_Tables_number           1
#_symmetry_space_group_name_H-M       "P11'"
_cell_length_a                        3.82781
_cell_length_b                        3.82781
_cell_length_c                        2.
_cell_angle_alpha                     9.
_cell_angle_beta                      9.
_cell_angle_gamma                     12.

loop_
_symmetry_equiv_pos_site_id
_symmetry_equiv_pos_as_xyz
   1    x,y,z

loop_
_atom_site_label
_atom_site_type_symbol
_atom_site_fract_x
_atom_site_fract_y
_atom_site_fract_z
_atom_site_occupancy
Si1 Si .33333 .66667 .48915 1.
Si2 Si .66666 .33333 .5185 1.

# Si_TDS.cif -- in P1.1 -- non-magnetic
```

```
data_Si_TDS_189
_audit_creation_date              222-8-4
_audit_creation_method            "Bilbao Crystallographic Server"
_symmetry_Int_Tables_number       1
#_symmetry_space_group_name_H-M   "P1'"
_cell_length_a                    6.43381
_cell_length_b                    6.43381
_cell_length_c                    2.
_cell_angle_alpha                 9.
_cell_angle_beta                  9.
_cell_angle_gamma                 12.

loop_
_symmetry_equiv_pos_site_id
_symmetry_equiv_pos_as_xyz
   1    x,y,z

loop_
_atom_site_label
_atom_site_type_symbol
_atom_site_fract_x
_atom_site_fract_y
_atom_site_fract_z
_atom_site_occupancy
Si1 Si .33333 .66667 .5 1.
Si2 Si .66666 .33333 .5 1.
Si3 Si .29982 . .5 1.
```

Si4 Si . .29982 .5 1.

Si5 Si .718 .718 .5 1.

Si6 Si . . .4322 1.

Si7 Si . . .56798 1.

# Si_HDS.cif -- in P1.1 -- non-magnetic

data_Si_HDS_189

| | |
|---|---|
| _audit_creation_date | 222-8-4 |
| _audit_creation_method | "Bilbao Crystallographic Server" |
| _symmetry_Int_Tables_number | 1 |
| #_symmetry_space_group_name_H-M | "P1'" |
| _cell_length_a | 6.29732 |
| _cell_length_b | 6.29732 |
| _cell_length_c | 2. |
| _cell_angle_alpha | 9. |
| _cell_angle_beta | 9. |
| _cell_angle_gamma | 12. |

loop_

_symmetry_equiv_pos_site_id

_symmetry_equiv_pos_as_xyz

   1    x,y,z

loop_

_atom_site_label

_atom_site_type_symbol

_atom_site_fract_x

_atom_site_fract_y

_atom_site_fract_z

_atom_site_occupancy

Si1 Si . . .5 1.

Si2 Si .35934 . .5 1.

Si3 Si . .35934 .5 1.

Si4 Si .6466 .6466 .5 1.

Si5 Si .33333 .66667 .56577 1.

Si6 Si .66666 .33333 .43423 1.

Si7 Si .33334 .66667 .43423 1.

Si8 Si .66667 .33334 .56577 1.

# Si_LHDS.cif -- in P1.1 -- non-magnetic

data_Si_LHDS_191

| | |
|---|---|
| _audit_creation_date | 222-8-4 |
| _audit_creation_method | "Bilbao Crystallographic Server" |
| _symmetry_Int_Tables_number | 1 |
| #_symmetry_space_group_name_H-M | "P11'" |
| _cell_length_a | 7.33381 |
| _cell_length_b | 7.33381 |
| _cell_length_c | 2. |
| _cell_angle_alpha | 9. |
| _cell_angle_beta | 9. |
| _cell_angle_gamma | 12. |

loop_

_symmetry_equiv_pos_site_id

```
_symmetry_equiv_pos_as_xyz
   1    x,y,z

loop_
_atom_site_label
_atom_site_type_symbol
_atom_site_fract_x
_atom_site_fract_y
_atom_site_fract_z
_atom_site_occupancy
Si1 Si .18192 .36384 .5 1.
Si2 Si .8188 .18192 .5 1.
Si3 Si .63616 .8188 .5 1.
Si4 Si .8188 .63616 .5 1.
Si5 Si .18192 .8188 .5 1.
Si6 Si .36384 .18192 .5 1.
Si7 Si .33333 .66667 .43292 1.
Si8 Si .66666 .33333 .43292 1.
Si9 Si .66666 .33333 .5678 1.
Si1 Si .33333 .66666 .5678 1.
```

## Acknowledgements

Additional assistance was granted through the computing cluster GRAFEN at Biocentrum Ochota, the Interdisciplinary Centre for Mathematical and Computational Modelling of Warsaw University (ICM UW) and Poznań Supercomputing and Networking Center (PSNC).

## Funding

This work was supported by the National Science Centre (NCN – Poland) Research Project: No. 2021/43/B/ST8/03207.